\documentclass[3p,times,twocolumn]{elsarticle}

\usepackage{ecrc}

\volume{00}

\firstpage{1}

\runauth{}

\jid{nuphbp}

\jnltitlelogo{Nuclear Physics B Proceedings Supplement}

\usepackage{amssymb}

\usepackage[figuresright]{rotating}

\begin{document}

\begin{frontmatter}




\vspace{-100 mm}

\title{Updated Results of a Solid-State Sensor Irradiation Study for ILC Extreme Forward Calorimetry \\ 
\bigskip Talk presented at the International Workshop on Future Linear Colliders (LCWS15), Whistler, Canada, 2-6 November 2015.}


\author{George Courcoubetis}
\author{Wyatt Crockett}
\author{Vitaliy Fadeyev}
\author{Caleb Fink}
\author{Nikolas Guillemaud}
\author{Cesar Gonzalez Renteria}
\author{Benjamin Gruey}
\author{Patrick LaBarre}
\author{Forest Martinez-McKinney}
\author{Greg Rischbieter}
\author{Bruce A. Schumm \corref{cor1}}
\ead{baschumm@ucsc.edu}
\author{Edwin Spencer}
\author{Max Wilder}
\cortext[cor1]{Corresponding author.}
\address{Santa Cruz Institute for Particle Physics and the University Of California, 1156 High Street,
Santa Cruz California 95064 USA}

\begin{abstract}
Detectors proposed for the International Linear Collider (ILC)
incorporate a tungsten sampling calorimeter (`BeamCal') intended to
reconstruct showers of electrons, positrons and photons
that emerge from the interaction point of the collider
with angles between 5 and 50 milliradians. For the
innermost radius of this calorimeter, radiation doses
at shower max are expected to reach 100 Mrad per year,
primarily due to minimum-ionizing electrons and positrons
that arise in the induced electromagnetic showers
of e$^+$e$^-$`beamstrahlung' pairs produced in the ILC beam-beam interaction. However,
radiation damage to calorimeter sensors may be dominated
by hadrons induced by nuclear interactions of shower photons,
which are much more likely to contribute to the non-ionizing
energy loss that has been observed to damage sensors exposed to
hadronic radiation. We report here on the results of SLAC
Experiment T-506, for which several different types of
silicon diode and gallium-arsenide sensors were exposed to doses of radiation
induced by showering electrons of energy 3.5-13.3 GeV. By embedding
the sensor under irradiation within a tungsten radiator, the exposure
incorporated hadronic species that would potentially contribute to the
degradation of a sensor mounted in a precision sampling calorimeter.
Depending on sensor technology, efficient charge collection
was observed for doses as large as 270 Mrad.

\end{abstract}


\end{frontmatter}


\section{Introduction}
\label{}

Far-forward calorimetry, covering the region between 5 and 50 milliradians
from the on-energy beam axis,
is envisioned as a component of both the ILD~\cite{ref:ILD_DBD} and SiD~\cite{ref:SiD_DBD}
detector concepts for the proposed International Linear Collider (ILC). The
BeamCal tungsten sampling calorimeter proposed to cover this angular region
is expected to absorb approximately 10 TeV of electromagnetic radiation
per beam crossing from e$^+$e$^-$ beamstrahlung pairs, leading to expected annual radiation doses of 100 Mrad
for the most heavily-irradiated portions of the instrument.
While the deposited energy is expected to arise primarily from minimum-ionizing
electrons and positrons in the induced electromagnetic showers,
radiation damage to calorimeter sensors may be dominated
by hadrons induced by nuclear interactions of shower photons,
which are much more likely to contribute to the non-ionizing
energy loss that has been observed to damage sensors exposed to
hadronic radiation. We report here on the latest results of SLAC
Experiment T-506, for which several different types of
silicon diode and gallium-arsenide (GaAs) sensors were exposed to doses of up to 270 Mrad
at the approximate maxima of electromagnetic
showers induced in a tungsten radiator by electrons of energy
3.5-13.3 GeV, similar to that of electrons and positrons
from ILC beamstrahlung pairs.

Bulk damage leading to the suppression of the electron/hole
charge-collection efficiency (CCE) is generally thought to be proportional
to the non-ionizing energy loss (`NIEL') component of the energy
deposited by the incident radiation.
Early studies of electromagnetically-induced damage to
solar cells~\cite{ref:TaeSung5,ref:TaeSung6,ref:TaeSung7}
suggested that p-type bulk sensors were more tolerant to
damage from electromagnetic sources, due to an apparent
departure from NIEL scaling, particularly for electromagnetic particles
of lower incident energy.

Several more-recent studies have
explored radiation tolerance of silicon diode to incident fluxes of electrons.
A study assessing the capacitance vs. bias voltage (CV) characteristics of sensors exposed
to as much as 1 Grad of incident 2 MeV electrons~\cite{ref:Rafi_09}
suggested approximately 35 times less damage to n-type magnetic
Czochralski sensors than that expected from NIEL scaling.
A study of various n-type sensor types exposed to 900 MeV electrons
showed charge-collection loss of as little as 3\% for exposures up to
50 Mrad exposure~\cite{ref:Dittongo2004}; for exposures of 150 Mrad, a suppression of damage
relative to NIEL expectations of up to a factor of four was observed~\cite{ref:Dittongo2005}.
These discrepancies have been attributed to the different types of
defects created by lattice interactions: electrons tend to create point-like defects that are more
benign than the clusters formed due to hadronic interactions.

Finally, in studies of sensors exposed to large doses of hadron-induced
radiation, p-type bulk silicon was found to be more radiation-tolerant
than n-type bulk silicon, an observation that has been attributed to the absence of type inversion and the
collection of an electron-based signal~\cite{ref:TaeSung3,ref:TaeSung4}.
However, n-type bulk devices have certain advantages, such as a natural inter-electrode
isolation with commonly used passivation materials such as silicon oxide and silicon nitride.

More recently, GaAs sensors have been proposed as a possible
radiation-tolerant alternative to silicon sensors, the latter of which can develop
large dark current after significant irradiation.
GaAs sensors, on the other hand, have been observed~\cite{ref:GaAs} to suffer significant
loss of CCE for moderate doses of 10-MeV-scale electrons.

Here, we report on an exploration of the radiation tolerance of silicon and GaAs sensors,
assessed via direct measurements of the median collected charge deposited
by minimum-ionizing particles. For some of the sensors, leakage current measurement
were also made.
Four different silicon-diode bulk compositions were explored: p-type and n-type doping of
both magnetic Czochralski and float-zone crystals. Both strip and pad geometries were
explored, depending on which sensor samples were available.
For strip sensors, p-type float-zone sensors were produced by Hamamatsu Photonics while the
remaining types were produced by Micron Corporation.
Sensor strip pitch varied between
50 and 100 $\mu$m, while the bulk thickness
varied between 307 $\mu$m (for the p-type magnetic Czochralski sensors)
and 320 $\mu$m (for the p-type float zone sensors).
A p-type float-zone pad sensor (part number WSI-P4), 320 $\mu$m thick and
manufactured as a test structure 
on the same wafer as ATLAS upgrade prototype sensors~\cite{ref:nobu}, was also exposed and studied.
Also explored was the radiation tolerance of 300 $\mu$-thick GaAs pad sensors produced
by means of the Liquid Encapsulated Czochralski method doped by a shallow donor 
(Sn or Te; Sn was used for the sensor in this study)~\cite{ref:GaAs}. Relative to a prior report~\cite{ref:T506_NIM} on the
radiation tolerance of silicon diode sensors, this report includes
new results on a 270 Mrad exposure and annealing study of the WSI-P4 p-type bulk
float-zone silicon diode pad sensor, as
well as new results based on a recent run of experiment T-506 in
which GaAs pad sensors was exposed to doses of 5.7 and 20.8 Mrad.

While the radiation dose was initiated by electromagnetic processes
(electrons showering in tungsten), the placement of the sensors near
shower max ensures that the shower incorporates an appropriate
component of hadronic irradiation arising from neutron spallation,
photoproduction, and the excitation of the $\Delta$ resonance.
Particularly for the case that NIEL scaling suppresses
electromagnetically-induced radiation damage, the small
hadronic component of the electromagnetic
shower might dominate the rate of damage to the sensor.
However, the size and effect of this component is difficult to
estimate reliably, and so we choose to study radiation damage in
a configuration that naturally incorporates all components present in
an electromagnetic shower.

\section{Experimental Setup}

Un-irradiated sensors were subjected to current vs. bias voltage (IV) and 
capacitance vs. bias voltage (CV) tests,
the results of which allowed
a subset of them to be selected for irradiation based on their
breakdown voltage (typically above 1000 V for selected sensors) and low level of leakage
current. The sensors were placed on carrier printed-circuit `daughter boards' and wire-bonded to a
readout connector. The material of the daughter boards was milled away in the
region to be irradiated in order to facilitate the
charge collection measurement (described below) and minimize radio-activation.
The median collected charge was measured with the Santa Cruz Institute for Particle Physics (SCIPP)
charge-collection (CC) apparatus (also described below) before irradiation.
The sensors remained mounted to their individual daughter boards throughout irradiation and the followup
tests, simplifying their handling and reducing uncontrolled annealing.
Additionally, this allowed a reverse-bias voltage to be maintained across the sensor during irradiation.
This voltage was kept small (at the level of a few volts) to avoid possible damage of the devices
from a large instantaneous charge during the spill.

Sensors were irradiated with beam provided by the End Station
Test Beam (ESTB) facility at the SLAC National Accelerator Laboratory.
Parameters of the beam provided by the ESTB facility are shown in
Tab.~\ref{tab:ESTB}. The beam was incident upon a series of
tungsten radiators.
An initial 7mm-thick (2.0 radiation-length) tungsten plate (`R1')
served to initiate the electromagnetic shower.
The small number of
radiation lengths of this initial radiator permitted the
development of a small amount of divergence of the shower
relative to the straight-ahead beam direction without
significant development of the largely isotropic hadronic
component of the shower.

\begin{table}[h]
\begin{centering}
\caption{Parameters of the beam delivered by the
ESTB facility during the T-506 experiment.}
\label{tab:ESTB}
\vspace {5mm}
\begin{tabular}{cc}
Parameter  &  Value   \\ \hline
Energy     &  3.5-13.3 GeV \\
Repetition Rate  &  5-10 Hz \\
Charge per Pulse  &  150-180 pC \\
Spot Size (radius) & $\sim 1$ mm \\
\end{tabular}
\par\end{centering}
\end{table}

This plate was followed by an
open length of approximately half a meter, which allowed a degree of
spreading of the shower before it impinged upon a second,
significantly thicker `R2' (4.0 radiation-length) tungsten plate,
which was followed immediately by the sensor undergoing
irradiation. This was closely followed, in turn, by an
8.0 radiation-length tungsten plate. Immediately surrounding the
sensor by tungsten radiators that both
initiated and absorbed the great majority of the electromagnetic
shower ensured that the sensor would be illuminated by a
flux of hadrons commensurate with that experienced by a calorimeter
sensor close to the maximum of a tungsten-induced shower.
More precise values of the location of the various radiator elements
and sensor, for each of the three years of running of T-506, are given in Tab.~\ref{tab:radiator}.

\begin{table}[h]
\begin{centering}
\caption{Location of the various radiator elements and 
the sensor under irradiation, for the three successive 
phases of T-506 running. The R1 radiator had a thickness
of 2 $X_0$, while the thickness of the R2 radiator was 4 $X_0$.
The apparent increased geometrical thickness of R2 in Year 1
was due to the presence of a 6mm air gap mid-way through the
radiator.}
\label{tab:radiator}
\vspace {5mm}
\begin{tabular}{lccc}
                &  Year 1        & Year 2        & Year 3   \\
                &  (2013)        &  (2014)       &  (2015)     \\
Surface         &  Location      & Location      & Location    \\ 
                &   (cm)         &    (cm)       & (cm)        \\
\hline
R1 Entrance  &  0.0  & 0.0 & 0.0  \\
R1 Exit      &  0.7  & 0.7 & 0.7  \\
R2 Entrance  &  55.0 & 45.7 & 46.6  \\
R2 Exit      &  57.0 & 47.1 & 48.0 \\
Sensor       &  57.7 & 47.6 & 48.5 \\
\end{tabular}
\par\end{centering}
\end{table}

Although initiating the shower significantly upstream of the sensor
promoted a more even illumination of the sensor
than would otherwise have been achieved, the half-width
of the resulting electron-positron fluence distribution
at the sensor plane was less than 0.5 cm. On the other hand,
the aperture of the CC apparatus (to be
described below) was of order 0.7 cm.. Thus, in order to
ensure that the radiation dose was well understood over
the region of exposure to the CC apparatus source,
it was necessary to achieve a uniform illumination over
a region of approximately 1 cm$^2$. This was done by
`rastering' the detector across the beam spot through
a range of 1 cm in both dimensions
transverse to that of the incident beam.
According to Monte Carlo simulation studies, this is expected
to generate a region of approximately 1 cm$^2$ over which
the illumination is uniform to within $\pm 20$\%. To account
for potential millimeter-level misalginments of the beamline
center with the sensor, a `targeting factor' of $(90 \pm 10$)\%
is included in the final dose-rate calculations.

\section{Dose Rates}

During the 120 Hz operation of the SLAC Linear Collider Light Source (LCLS),
5-10 Hz of beam was deflected by a pulsed kicker magnet into the End Station transfer line.
The LCLS beam was very stable with respect to both current and energy. Electronic
pickups and ion chambers measured the beam current and beam loss through the
transfer line aperture, ensuring that good transfer efficiency could be established
and maintained. The transfer efficiency was estimated to be ($95 \pm 5$)\%. 

To calculate the dose rate through the sensor, it is necessary to determine
the `shower conversion factor' $\alpha$ that provides the mean fluence of minimum-ionizing
particles (predominantly electrons and positrons), in particles per cm$^2$,
per incoming beam electron. This factor is dependent upon the radiator
configuration and incident beam energy, as well as the rastering pattern
used to provide an even fluence across the sensor (as stated above,
the detector was translated continuously across the beam centerline
in a 1 cm$^2$ square pattern).

To estimate $\alpha$, the Electron-Gamma-Shower (EGS) Monte Carlo program~\cite{ref:EGS}
was used to simulate showers through the radiator configuration
and into the sensor. The radiator configuration
was input to the EGS program, and a mean fluence profile (particles per
cm$^2$ through the sensor as a function of transverse distance from the nominal
beam trajectory) was accumulated by simulating the showers of 1000
incident electrons of a given energy. To simulate the rastering process,
the center of the simulated profile was then
moved across the face of the sensor in 0.5mm steps, and an estimated mean fluence
per incident electron as a function of position on the sensor (again, relative to the nominal beam
trajectory) was calculated. This resulted in a mean fluence per incident electron
that was uniform to within 20\% 1mm anywhere inside the boundary of the rastering region.
The value of $\alpha$ used for subsequent irradiation dose estimates was taken to be
the value found at the intersection of the nominal
beam trajectory with the sensor plane. The simulation was repeated for
various values of the incident electron energy, producing the values of
$\alpha$ shown in Tab.~\ref{tab:alpha_2013} (Tab.~\ref{tab:alpha_2014-15})
for the 2013 (2014-15) radiator configuration.
For years 2014 ans 2015, the spacings
of the radiator and sensor were close enough that
a single mean value of $\alpha$ sufficed.

\begin{table}[h]
\begin{centering}
\caption{Shower conversion factor $\alpha$, giving
the mean fluence at the sensor per incident
electron, as a function of electron energy. for the
2013 radiator configuration. These
values include the effect of rastering over a 1 cm$^2$
area surrounding the nominal beam trajectory.
Also shown is the number of rads per nC of delivered
charge, at the given energy, corresponding to the
given value of $\alpha$. 
}
\label{tab:alpha_2013}
\vspace {5mm}
\begin{tabular}{ccc}
Beam      & 2013 Shower                     & Dose per nC             \\
Energy    &   Conversion                    & Delivered                \\
(GeV)     & Factor $\alpha$                 & Charge (krad)           \\ \hline
2  &  2.1 & 0.34  \\
4  &  9.4 & 1.50  \\
6  & 16.5 & 2.64  \\
8  & 23.5 & 3.76  \\
10 & 30.2 & 4.83  \\
12 & 36.8 & 5.89  \\
\end{tabular}
\par\end{centering}
\end{table}

\begin{table}[h]
\begin{centering}
\caption{Shower conversion factor $\alpha$, giving
the mean fluence at the sensor per incident
electron, as a function of electron energy, 
for the 2014-15 radiator configuration. These
values include the effect of rastering over a 1 cm$^2$
area surrounding the nominal beam trajectory.
Also shown is the number of rads per nC of delivered
charge, at the given energy, corresponding to the
given value of $\alpha$. For 2014 and 2015, the spacings
of the radiator and sensor were close enough that
a single mean value of $\alpha$ sufficed.
}
\label{tab:alpha_2014-15}
\vspace {5mm}
\begin{tabular}{ccc}
Beam      & 2014-15 Shower      & Dose per nC             \\
Energy    & Conversion          & Delivered                \\
(GeV)     & Factor $\alpha$     & Charge (krad)           \\ \hline
3  & 4.6 &  0.73  \\
5  & 10.0 & 1.60  \\
7  & 15.5 & 2.48  \\
9  & 21.1 & 3.38  \\
11 & 26.7 & 4.27  \\
13 & 31.8 & 5.09  \\
15 & 37.7 & 6.03  \\
17 & 43.0 & 6.88  \\
\end{tabular}
\par\end{centering}
\end{table}

To convert this number to rads per nC of delivered charge, a mean
energy loss in silicon of 3.7 MeV/cm was assumed, leading to
a fluence-to-rad conversion factor of 160 rad per nC/cm$^2$.
It should be noted that while this dose rate considers only
the contribution from electrons and positrons, these
two sources dominate the overall energy absorbed by the
sensor. In addition, the BeamCal dose-rate spec of 100 Mrad
per year considered only the contribution from electrons
and positrons.

To confirm the adequacy of the dose-calibration simulation, 
in 2013 an
in-situ measurement of the dose was made using a
radiation-sensing field-effect transistor (`RADFET')~\cite{ref:radfet}
positioned on a daughter board at the expected
position of the nominal beam trajectory at the
center of the rastering pattern.
Beam was delivered in 150 pC pulses of 4.02 GeV
electrons; a total of 1160 pulses were directed
into the target over a period of four minutes,
during which the sensor was rastered quickly
through its 1 cm$^2$ pattern.
The RADFET was then read out, indicating
a total accumulated dose of 230 krad,
with an uncertainty of roughly 10\%. Making
use of the dose rate calibration of Tab.~\ref{tab:alpha_2013},
interpolating to the exact incident energy of 4.02 GeV,
and taking into account the ($95 \pm 5$)\% transfer efficiency
of the ESTB beamline, leads to an expected dose of 250 krad,
within the $\sim$10\% uncertainty of the RADFET measurement.

\section{Sensor Irradiation Levels}

Four types of silicon diode sensors were studied:
p-type and n-type doped versions of
both magnetic Czochralski and float-zone crystals.
In what follows, we will use the notation `N' (`P')
for n-type (p-type) bulk sensors, and `F' (`C')
for float-zone (magnetic Czochralski) crystal technology.
In addition, two GaAs sensors were irradiated.
Once a sensor was irradiated with the ESTB, it was placed
in a sub-freezing environment and not irradiated again.
Up to four sensors of each type were irradiated and
chilled until they could be brought back to the University
of California, Santa Cruz campus for the
post-irradiation CC and leakage current measurements. In
addition, the sub-freezing environment was maintained
both during and after the measurements, so
that controlled annealing studies can eventually be done.

Table~\ref{tab:dose} displays the dose parameters of the
irradiated sensors. The
$(95 \pm 5)$\% transfer line efficiency 
and the $(90 \pm 10)$\% targeting factor have been taken
into account in these estimates. The numeral following
the two letters in the sensor identifier refer to
an arbitrary ordering of sensors assigned during
the sensor selection. Sensors were held at between 0 and 5 C
during irradiation. With the exception of sensor
NC02, which was accidentally annealed for 5 hours at temperatures as high as
130 C, all sensors were transferred to a cold
(below -10 C) environment immediately after irradiation.
All four silicon diode sensor types were exposed to dose rates
of approximately 5 and 20 Mrad, while an NF sensor
received over 90 Mrad, an NC sensor 220 Mrad and a PF pad sensor 270 Mrad.
Two GaAs sensors received doses of approximately 6 and 21 Mrad, respectively.
CC and leakage current results for the
irradiated sensors will be presented below.

\begin{table*}[h]
\begin{centering}
\caption{Dose parameters of the irradiated sensors. A
$(95 \pm 5)$\% transfer line efficiency and
a $(90 \pm 10)$\% targeting factor has been taken
into account in final dose estimates. While the NC02
sensor was irradiated at a temperature of 5 C,
it was accidentally annealed for approximately 5 hours
at temperatures as high as 130 C.}
\label{tab:dose}
\vspace {5mm}
\begin{tabular}{lccccc}
Sensor  & Strip (S)  & Year &  Beam Energy          &   Delivered      &  Dose     \\
        & or Pad (P) & 2013 & (GeV)                &  Charge ($\mu$C) & (Mrad)    \\ \hline \hline
PF05    & S          & 2013 &  5.88                & 2.00             & 5.1   \\
PF14    & S          & 2013 &  3.48                & 16.4             & 19.7  \\ \hline
PC10    & S          & 2013 &  5.88                & 1.99             & 5.1   \\
PC08    & S          & 2013 &  (5.88, 4.11, 4.18)  & (3.82,3.33,3.29) & 20.3  \\ \hline
NF01    & S          & 2013 &  4.18                & 2.30             & 3.7   \\
NF02    & S          & 2013 &  4.02                & 12.6             & 19.0  \\
NF07    & S          & 2013 &  8.20                & 23.6             & 91.4  \\ \hline
NC01    & S          & 2013 &  5.88                & 2.00             & 5.1   \\
NC10    & S          & 2013 &  3.48                & 15.1             & 18.0  \\
NC03    & S          & 2013 &  4.01                & 59.9             & 90.2  \\
NC02    & S          & 2013 &  (10.60,8.20)        & (32.3,13.8)      & 220   \\  \hline
GaAs-18 & P          & 2014 &  3.87                &  6.03            & 5.7   \\ 
GaAs-09 & P          & 2014 &  3.90                &  21.7            & 20.8  \\  \hline
WSI-P4 (PF)  & P     & 2015 &  13.32               &  50.9            & 269   \\  \hline           
\end{tabular}
\par\end{centering}
\end{table*}

\section{Charge Collection Measurement}

The SCIPP CC
apparatus incorporates a $^{90}$Sr source that has a secondary $\beta$-decay
with an end-point energy of 2.28 MeV. These $\beta$ particles illuminate
the sensor under study, which is read out in one of two ways,
depending upon whether it is a strip or pad sensor. These two 
charge collection readout approaches are described in detail below. 

For strip sensors, 64 channels are read out by the PMFE ASIC~\cite{ref:PMFE},
with a shaping time of 300 nsec. Whenever one of the 64 channels exceeds
a pre-set, adjustable threshold, the time and duration of the excursion
over threshold is recorded. In addition, the $\sim$20 Hz of $\beta$ particles that pass through
the sensor, and subsequently
enter a small (2mm horizontal by 7mm vertical) slit, trigger a
scintillator, and the time of excitation of the scintillator is also recorded.
If the slit is properly aligned with the read-out channels of
the sensor, and the sensor is efficient at the set read-out threshold,
a temporal coincidence between the scintillator pulse and
one of the read-out channels will be found in the data stream.

\begin{figure}[h]
  \begin{center}
    \includegraphics[width=0.45\textwidth]{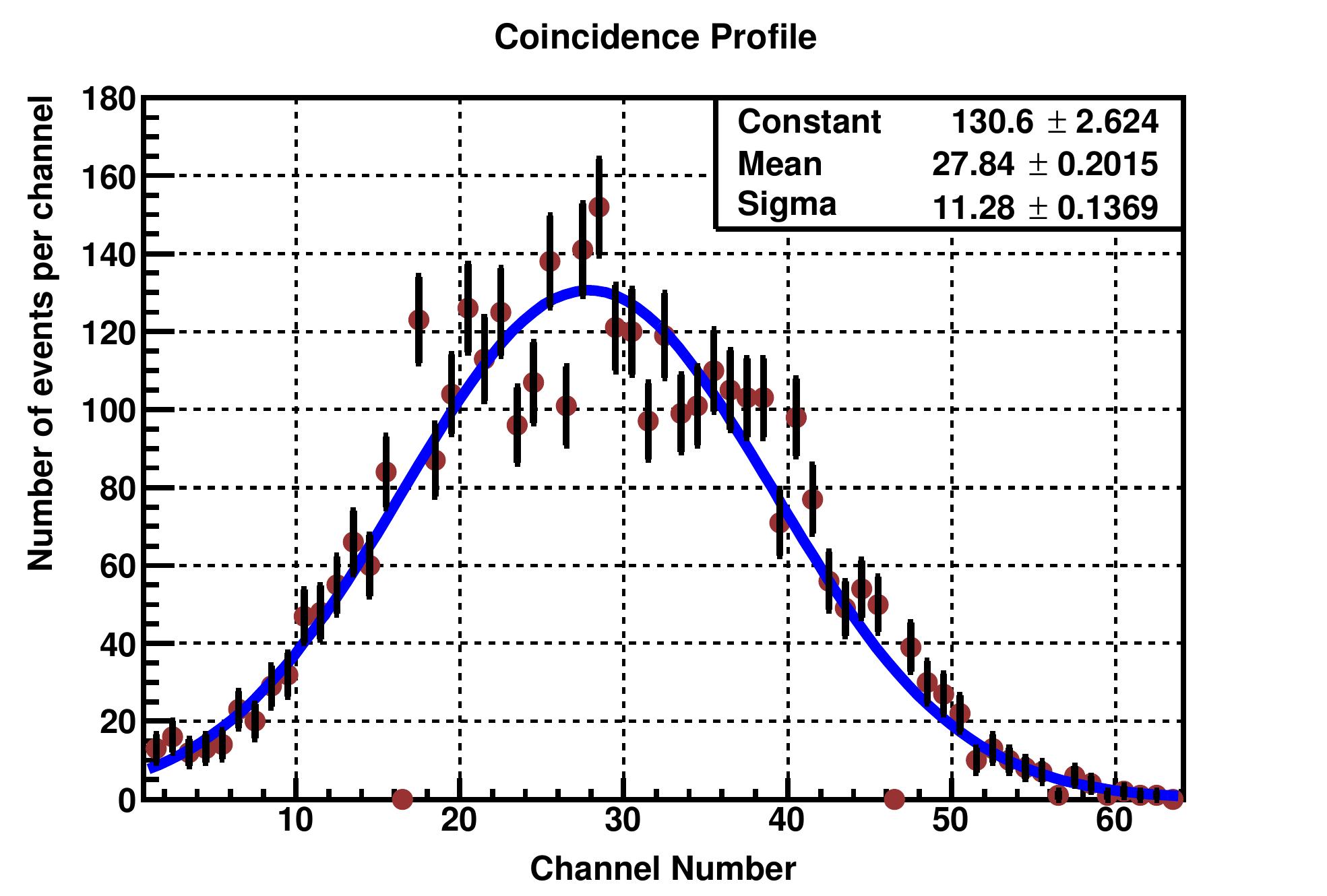}
  \end{center}
  \caption{
Sample profile of coincidences between the read-out
sensor channels and the trigger scintillator. The integral
of this distribution provides a count of the number of
$\beta$ particles triggering the scintillator that also
exceed the chosen PMFE threshold in one of the read-out channels.
\label{fig:coinc}
  }
\end{figure}

\begin{figure}[h]
 \begin{center}
   \includegraphics[width=0.45\textwidth]{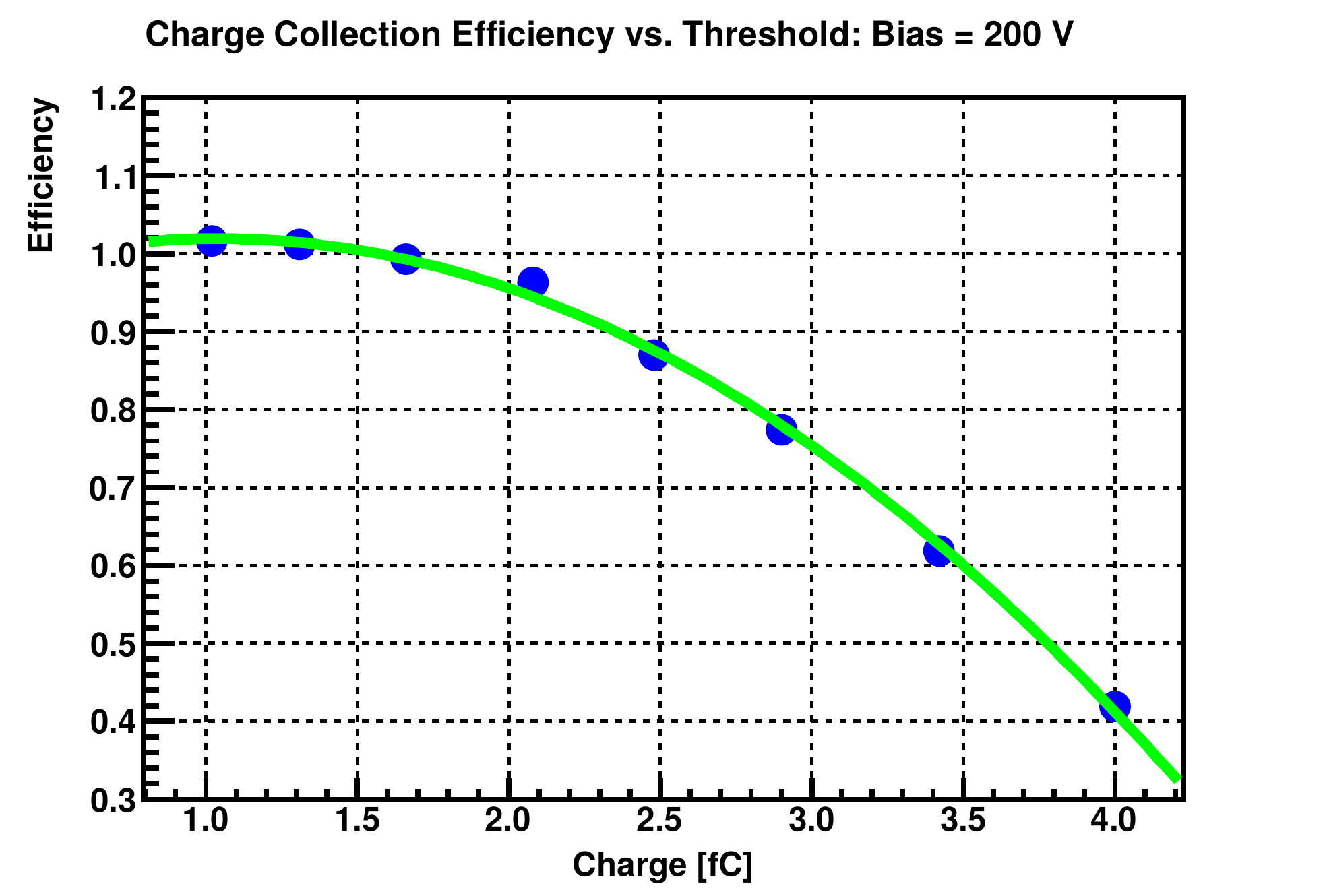}
 \end{center}
 \caption{
Plot of efficiency vs. PMFE threshold setting for
one of the irradiated sensors. The abscissa has
been converted from applied threshold voltage to
the amount of input PMFE charge that will exceed
the given threshold exactly 50\% of the time.
The point at which the curve
crosses the 50\% level
yields the median CC for the given bias level.
\label{fig:thresh_curve}
}
\end{figure}

Figure~\ref{fig:coinc} shows a sample coincidence profile
(histogram of the number of coincidences vs. channel
number) for a strip sensor for a 150-second run at a given threshold and
reverse bias level for one of the irradiated sensors
(specifically, for the NC01 sensor after 5.1 Mrad
of irradiation, applying a 300V reverse bias and a 130 mV threshold).
The integral of the distribution yields an estimate
of the total number of coincidences found during
the run, which, when divided by the number of scintillator
firings (after a small correction for cosmic background events)
yields the median CC level at that threshold
and bias level. 
This measurement can then be performed as
a function of threshold level, yielding the curve shown in
Fig.~\ref{fig:thresh_curve}. For this plot, the abscissa
has been converted from voltage (the applied threshold
level) to fC (the PMFE input charge that will fire the threshold
with exactly 50\% efficiency) via a prior calibration step
involving measurement of the PMFE response to known
values of injected charge. The point at which the curve
in Fig.~\ref{fig:thresh_curve} crosses the 50\% level
yields the median CC for the given bias level. 
In a prior study of sensors irradiated with hadrons, the SCIPP
apparatus gave median charge results consistent with that of
other charge collection systems used to assess radiation damage in
that study~\cite{Hara}.

For assessing the CCE of pad sensors, 
a two-stage, single-channel amplifier was constructed from discrete components,
based on a design of Fabris, Madden and Yaver~\cite{ref:jfet_amp}.
For the first stage, a cascode of two NXP BF862 JFETs is used. The source of the second 
JFET was connected to the non-inverting input of an LM6171 operational amplifier,
chosen for its high slew rate and low input noise contribution. 
The output of this opamp was then fed back to the input of the first JFET, 
completing the negative feedback loop. An external network, including 
a 32 dB Sonoma Instrument 310 SDI amplifier, was used to further amplify the pulse and 
shape it to a rise-time of 290 ns. 

\begin{figure}[h]
 \begin{center}
   \includegraphics[width=0.45\textwidth]{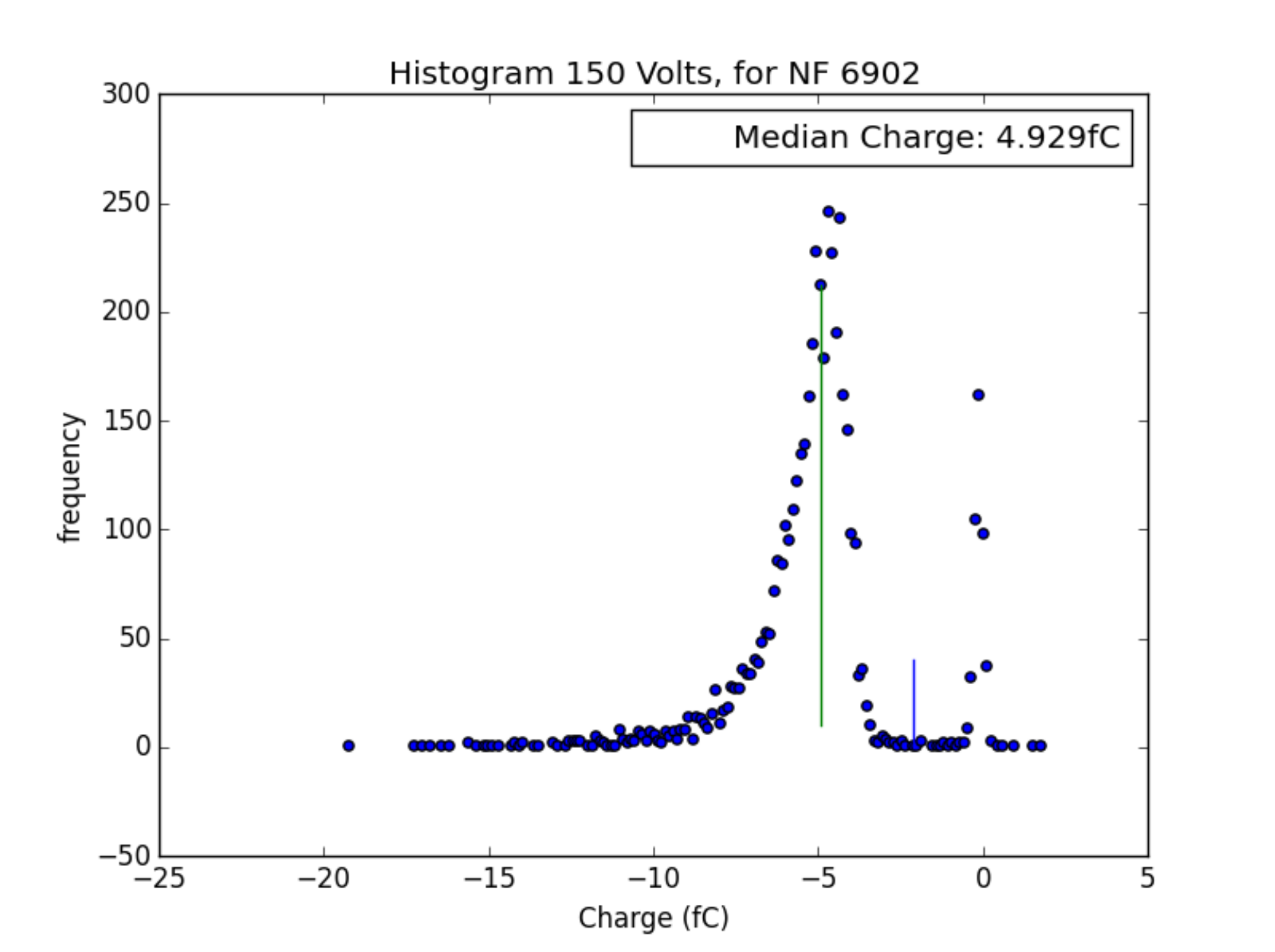}
 \end{center}
 \caption{
Histogram of pulse height for photomultiplier-triggered data events for
the single-channel readout. Both
the Landau distribution due to through-going $\beta$ particles as well
as the noise pedestal (for triggers for which the $\beta$ particle did not
traverse the sensor) are seen.
\label{fig:PH_dist}
}
\end{figure}

Upon receiving a trigger from the scintillator, the signal 
from the amplifier was read out by a Tektronix DPO 4054
digital storage oscilloscope, and the digitized waveforms
were written out and stored on the disk of a dedicated
data-acquisition computer. After the waveforms were accumulated on the computer,
a narrow temporal window was set around the peak of the average excitation pulse
from through-going beta particles, and a histogram was made of the resulting
pulse-height distribution; a typical distribution is shown in Fig.~\ref{fig:PH_dist}.
Since not all $\beta$ particles that trigger the scintillator go through the pad,
the distribution shows contributions from both the Landau deposition of the
through-going $\beta$ particles, as well as that of the noise pedestal, allowing
for an in-situ subtraction of the mean pedestal.

The amplification system was calibrated by reading out an unirradiated silicon diode
sensor of known thickness, and
comparing the median charge of the resulting Landau
distribution (after subtracting off the mean pedestal) to that expected
for an unirradiated sensor of that thickness.
This yields a gain of approximately 40 mV/fC, with a small dependence on load
capacitance. The width of the pedestal distribution then provides a measurement of the 
readout noise, which was found to be approximately 250 electrons at room temperature. 

\section{Charge Collection and Leakage Current Results}

The daughter boards
containing the irradiated sensors were designed
with connectors that allowed them to be attached to the
CC apparatus readout board without handling the sensors.
The median CC was measured as a function of reverse bias voltage for each sensor
both before and after irradiation.

\begin{figure}[h]
 \begin{center}
   \includegraphics[width=0.45\textwidth]{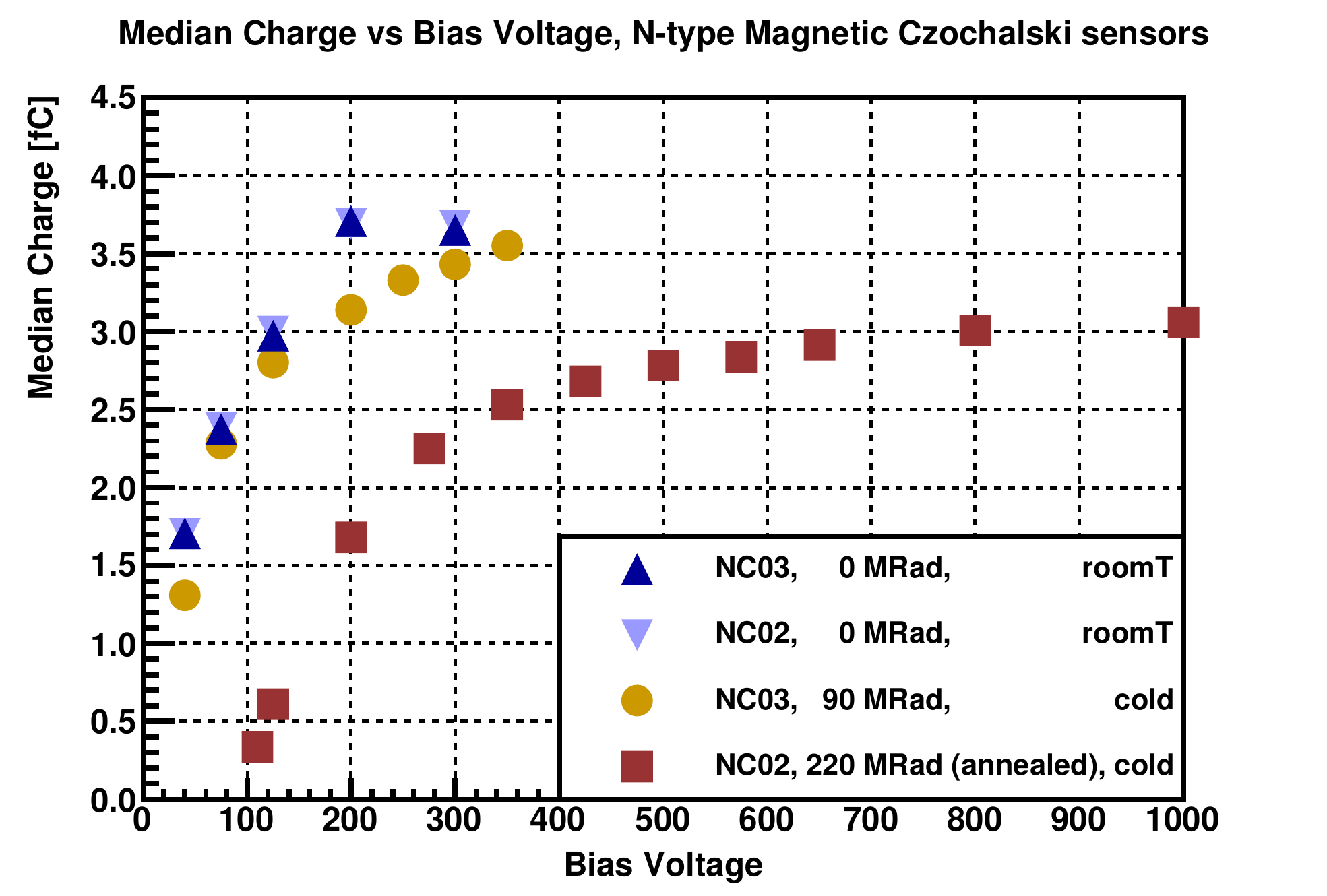}
 \end{center}
 \caption{
Median collected charge [fC] for the NC03 (90 Mrad dose) and NC02 (220 Mrad dose)
sensors.
\label{fig:NC_CCE}
}
\end{figure}

A series of irradiation runs of between 5 and 220 Mrad was taken 
for the NC (n-type bulk magnetic Czochralski) sensor type. For the exposures of 5.1 (NC01)
and 18.0 Mrad (NC10), no difference in CC performance was observed
relative to the pre-irradiation studies of the NC01 and NC10 sensors.
In Fig.~\ref{fig:NC_CCE} the median CC
both before and after irradiation is plotted for the NC03 (90 Mrad dose)
and NC02 (220 Mrad dose) sensors; it should be borne in mind,
though, that the NC02 sensor experienced significant annealing before
the post-irradiation measurement was done. It is seen that,
while the depletion voltage increases significantly with dose, median CC
within 20\% of un-irradiated values is maintained
for doses above 200 Mrad, although it may require annealing
to maintain efficiency at that level.

\begin{figure}[h]
 \begin{center}
   \includegraphics[width=0.50\textwidth]{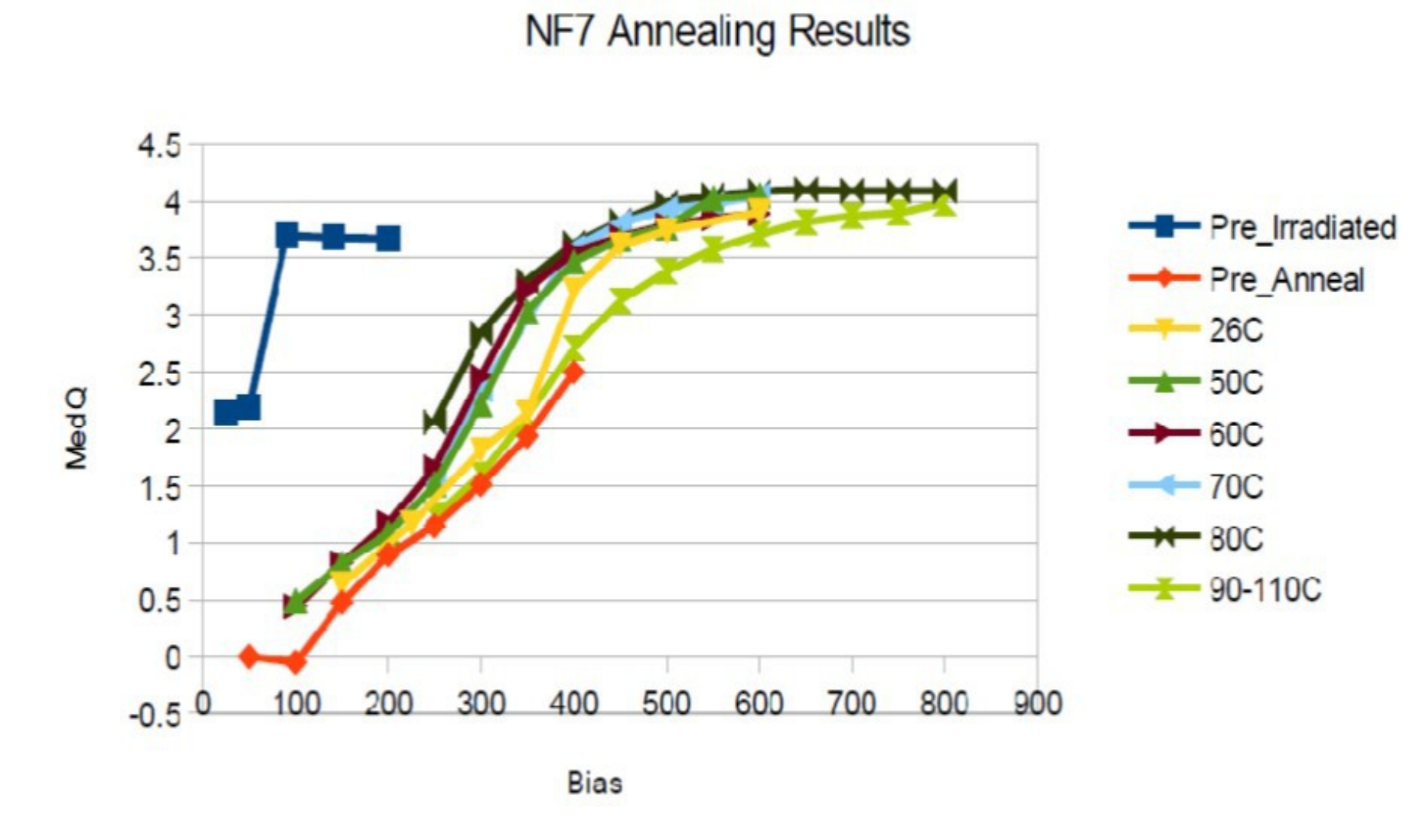}
 \end{center}
 \caption{
High-dose CC results for the NF07 (90 Mrad dose) sensor, as a function
of annealing temperature. The sensor was annealed at the reported temperature
for one hour.
\label{fig:NF_anneal}
}
\end{figure}

A comprehensive study of the effects of annealing on
CC was performed for an NF-type sensor after 
an exposure of 90 Mrad. 
Figure~\ref{fig:NF_anneal} exhibits the CC results
for this sensor after annealing
for one-hour periods over a series of progressively higher temperatures.
After a one-hour annealing step at room temperature, full CC
is observed for NF07 even after irradiation to nearly 100 Mrad, although
the depletion voltage rose significantly with irradiation. The
depletion voltage was observed to decrease significantly for
annealing temperatures below 80$^o$ C (beneficial annealing), but then rose again for
annealing temperatures in excess of 100$^o$ C (reverse annealing).

\begin{figure}[h]
 \begin{center}
   \includegraphics[width=0.50\textwidth]{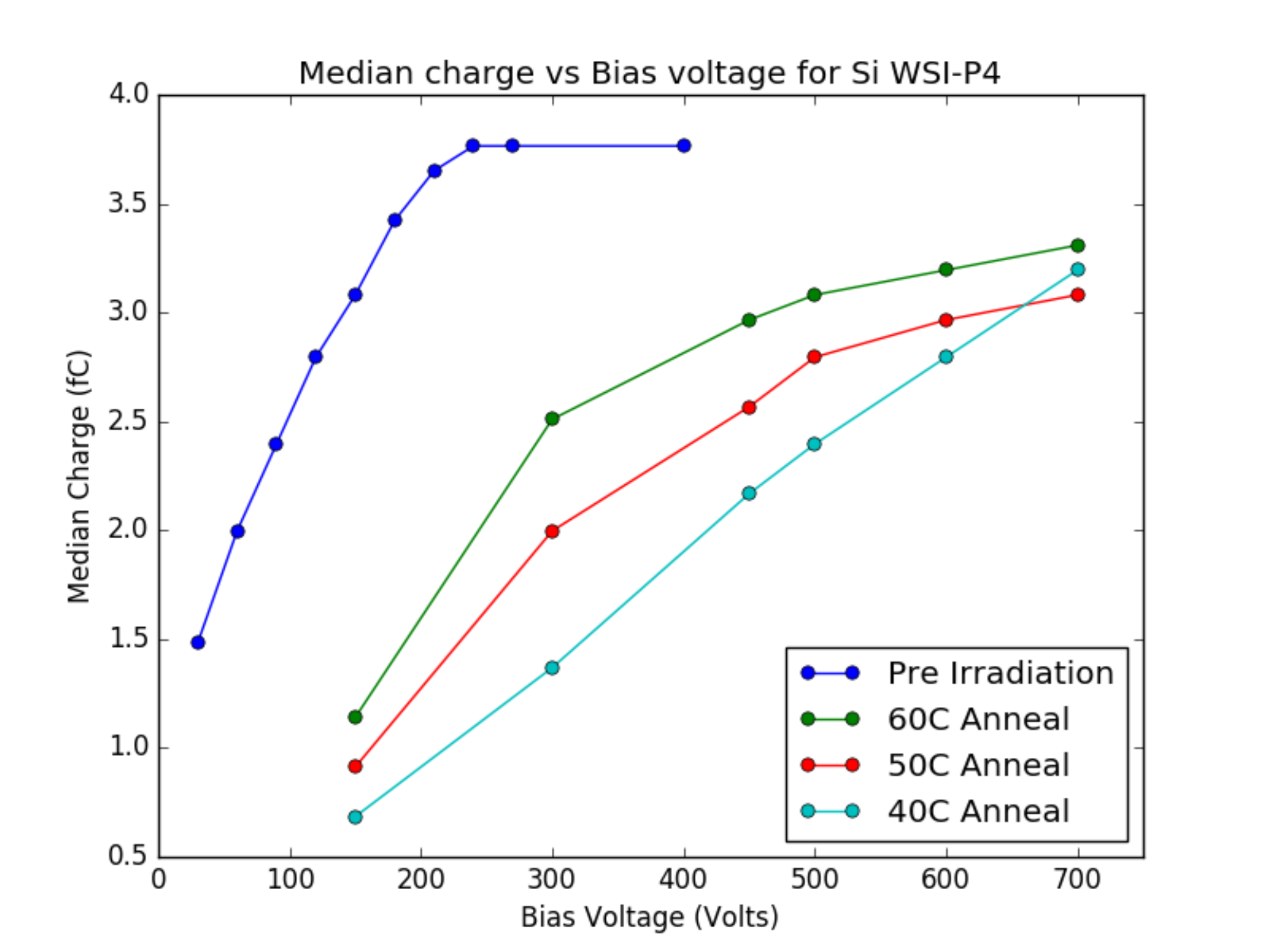}
 \end{center}
 \caption{
CC results for the WSI-P4 (PF pad) sensor before and after
irradiation with a 270 Mrad dose, and after a series of annealing
steps. Significant CCE remained after the 270 Mrad
dose, with both the depletion voltage and CCE 
showing improvement with annealing at temperature of up
to 60$^o$ C.
\label{fig:PF_300_CC}
}
\end{figure}

\begin{figure}[h]
 \begin{center}
   \includegraphics[width=0.50\textwidth]{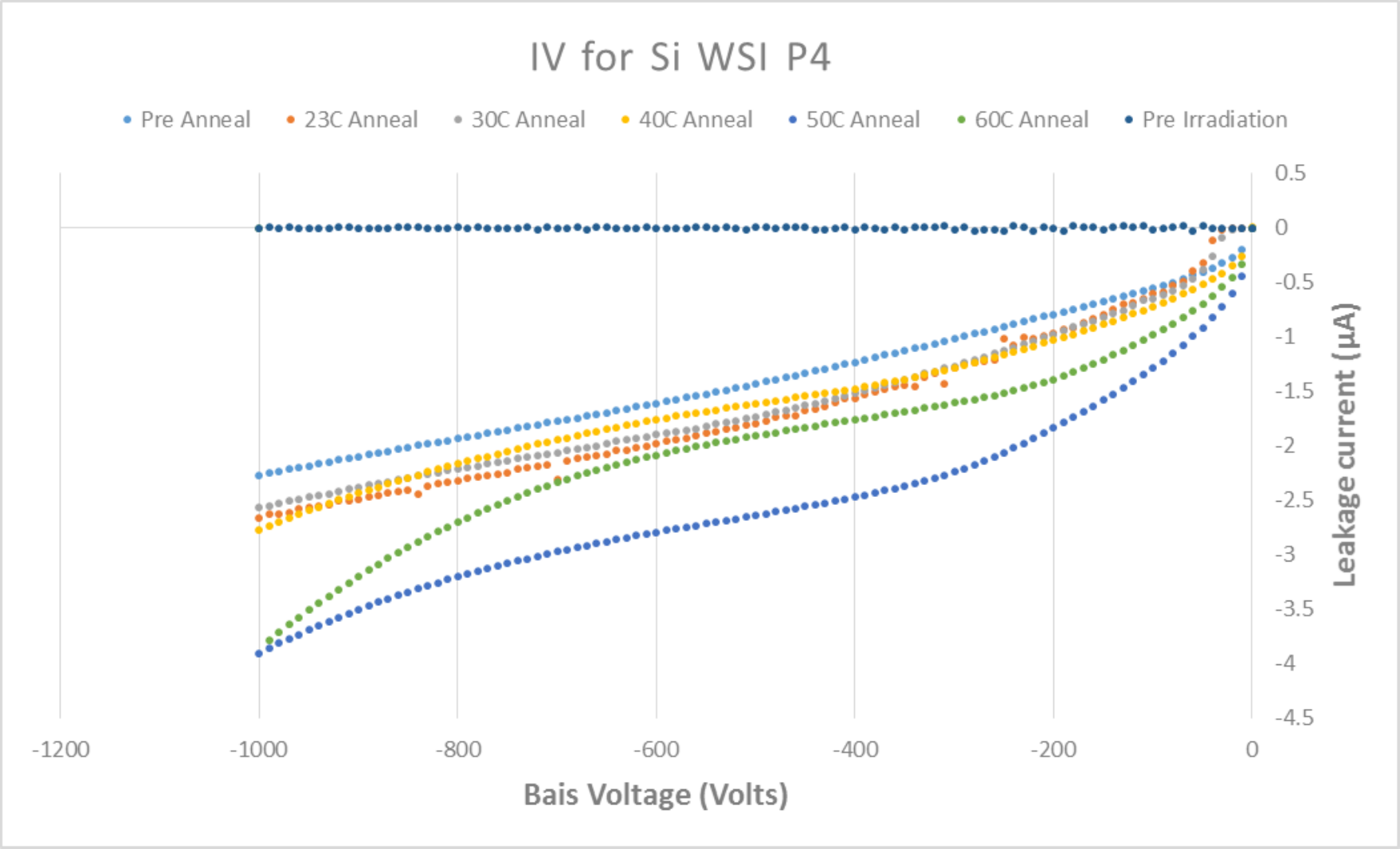}
 \end{center}
 \caption{
Leakage current  at -10$^o$ C for the WSI-P4 p-type float-zone pad sensor after
irradiation with a 270 Mrad dose, and after a series of annealing
steps. Pre-irradiation leakage current was measured to be below 10 nA.
\label{fig:PF_300_IV}
}
\end{figure}


In 2015, a p-type float-zone pad sensor (WSI-P4) was irradiated to a total dose of 270 Mrad,
and both leakage current and CCE were studied, both directly after
irradiation as well as after one-hour annealing periods at successively 
higher temperature. Figure~\ref{fig:PF_300_CC} shows the post-irradiation
CCE (for 300 ns shaping time) as a function of bias 
voltage for the various stages of annealing. The CCE
was seen to increase monotonically with bias voltage, reaching a maximum
of approximately 85\% of the pre-irradiation value above 600 V after a
one-hour annealing period at 60$^o$ C.   

Figure~\ref{fig:PF_300_IV} shows the post-irradiation leakage
current for sensor WSI-P4, measured at -10$^o$C, as a function of bias 
voltage for the various stages of annealing.
Relative to the pre-irradiation level of less than 10 nA, the current is seen to be significantly
larger after irradiation, with little improvement observed as the sensor
is annealed at successively higher temperature.
At a bias voltage of 600 V
the leakage current is observed to be about 2 $\mu$A, corresponding to a 
power draw of approximately 1.2 mW. Taking into account the 0.025 cm$^2$ area
of the sensor, this corresponds to a post-irradiation power draw of approximately 50 mW/cm$^2$
after an exposure of 270 Mrad.

\begin{figure}[h]
  \begin{center}
    \includegraphics[width=0.50\textwidth]{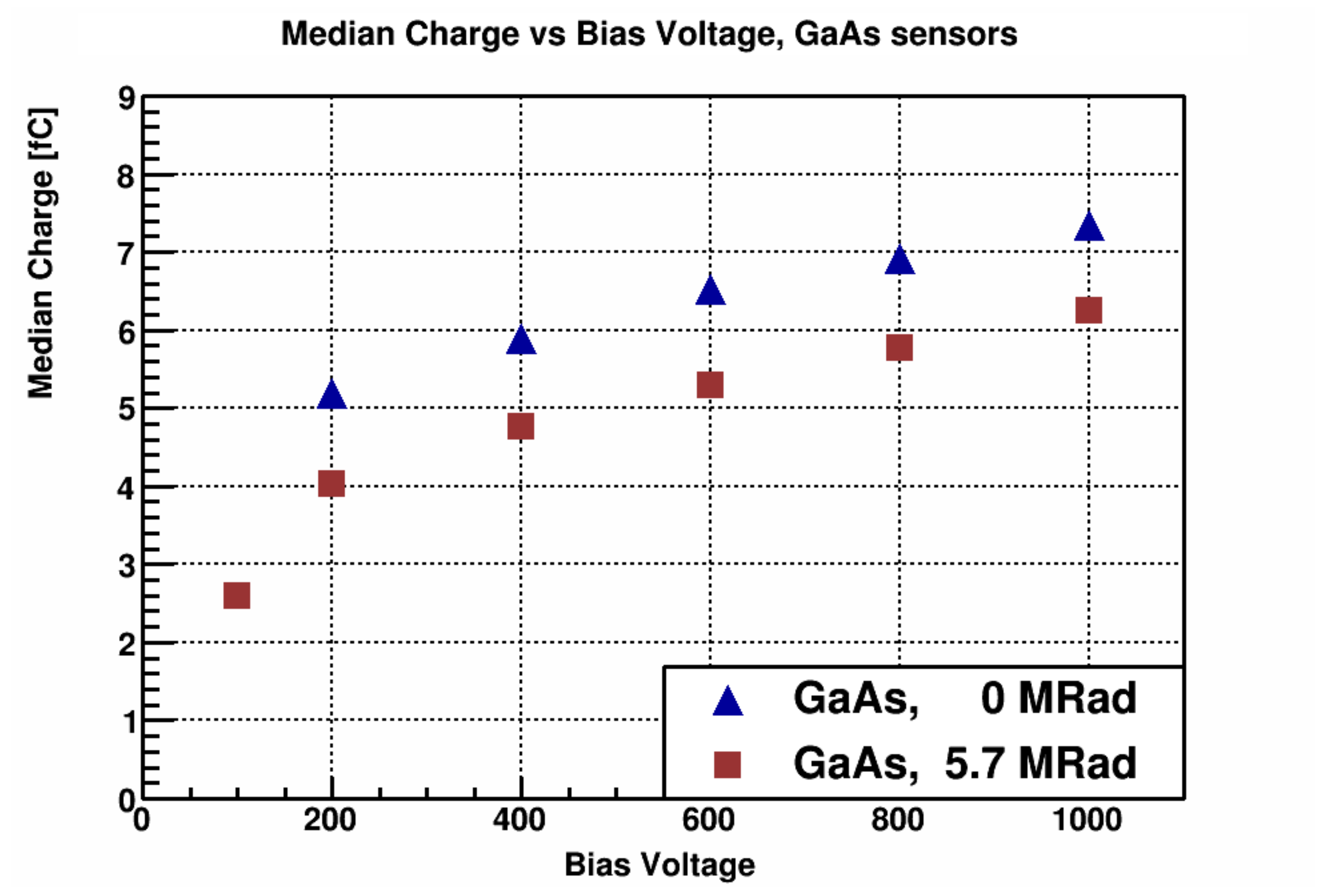}
  \end{center}
  \caption{
GaAs-18 sensor CCE before and after irradiation with a 5.7 Mrad dose.
    \label{fig:GaAs_6}
  }
\end{figure}

For the GaAs sensors, CCE and leakage current
were measured both before and
after exposures of 5.7 and 20.8 Mrad.
Figure~\ref{fig:GaAs_6} shows the CCE as a function of
bias voltage for the GaAs-18 sensor before and after
a 5.7 Mrad irradiation, and then after an annealing
step for which the sensor was
heated to 23$^o$ C for one hour.
Since the GaAs sensor is
not a diode, the CC does not reach a maximum as the
applied voltage is raised. However, the CC is observed
to drop by approximately 20\% over the full range of applied
voltage. This observation is roughly consistent with that
of~\cite{ref:GaAs}, which made use of an exposure of
a pure beam of order-10-MeV electrons. Room-temperature
annealing was seen to further degrade the CCE.

\begin{figure}[h]
 \begin{center}
   \includegraphics[width=0.50\textwidth]{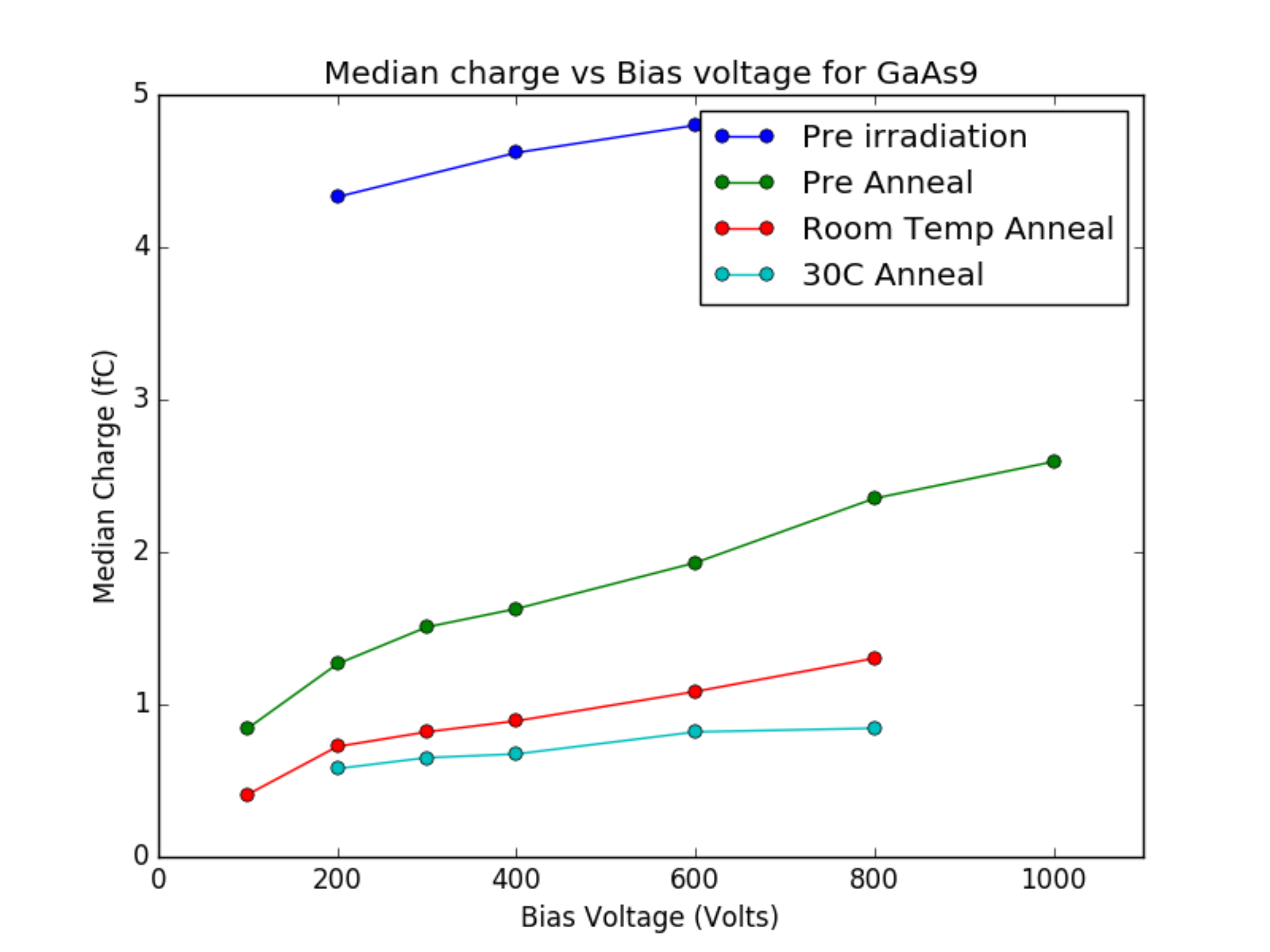}
 \end{center}
 \caption{
CCE results for the GaAs-09 sensor before and after
irradiation with a 20.8 Mrad dose, and after a series of annealing
steps. CCE was observed to degrade substantially
with irradiation, and degrade further with annealing. 
\label{fig:GAAS_21_CCE}
}
\end{figure}

Figure~\ref{fig:GAAS_21_CCE} shows the CCE as a function of
bias voltage for the GaAs-09 sensor before and after
a 20.8 Mrad irradiation, and then after an annealing
step at room temperature and at 30$^o$ C. Before annealing,
but after irradiation, the CCE at a bias of 600 V was seen to decrease
by over 40\%; after annealing at 30$^o$ C, the CCE was further
reduced to a level of approximately 15\% of its pre-irradiated
value. Again, this CCE loss after 21 Mrad of exposure is consistent with that
of~\cite{ref:GaAs}.

\begin{figure}[h]
 \begin{center}
   \includegraphics[width=0.50\textwidth]{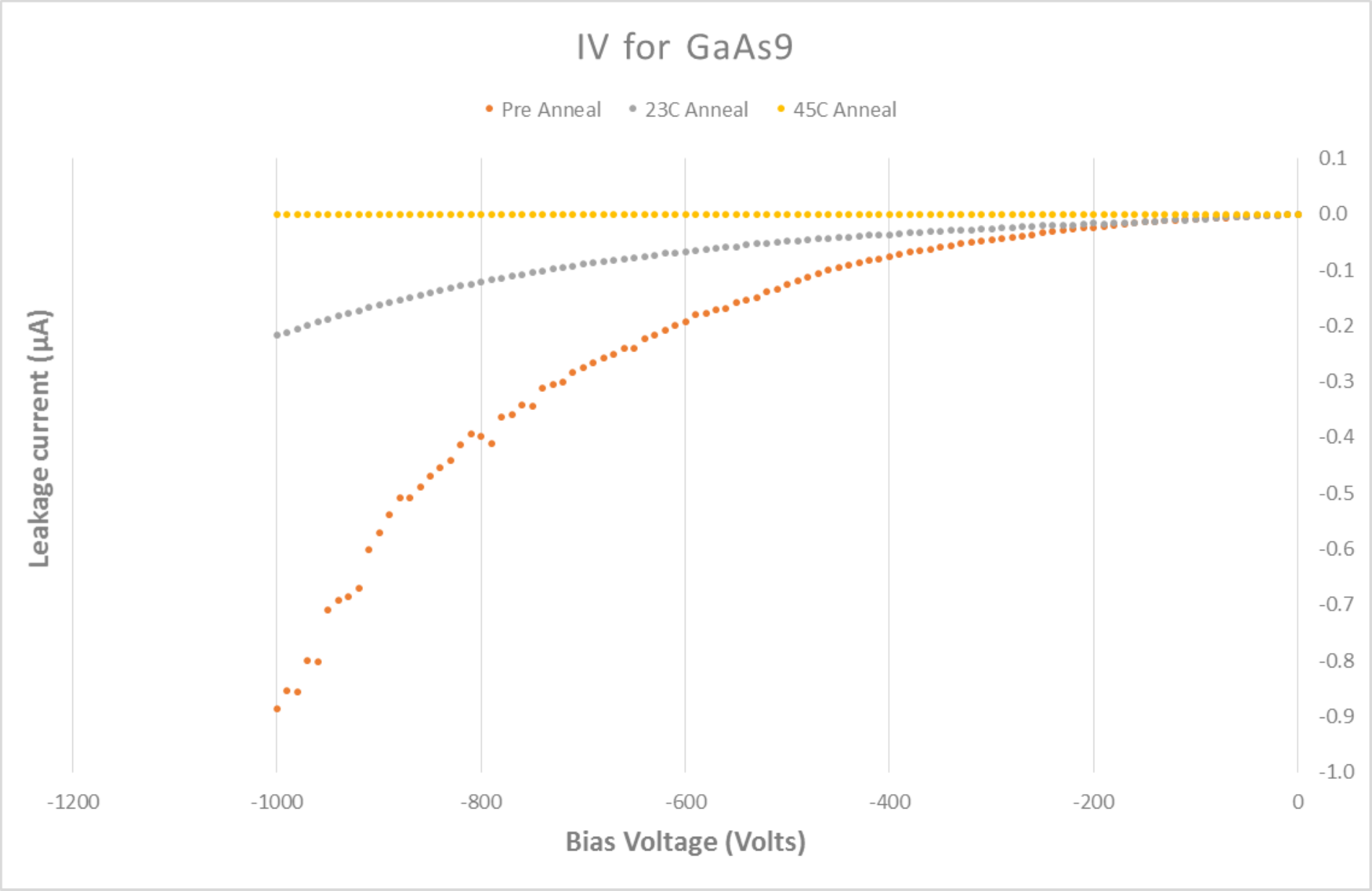}
 \end{center}
 \caption{
Leakage current at -10$^o$ C for the GaAs-09 sensor after
irradiation with a 20.8 Mrad dose, and after a series of annealing
steps. Pre-irradiation leakage current was measured to be below 10 nA.
\label{fig:GAAS_21_IV}
}
\end{figure}

Figure~\ref{fig:GAAS_21_IV} 
shows the post-irradiation leakage
current for GaAs-09 (20.8 mRad exposure), measured at -10$^o$C, 
as a function of bias voltage for the various stages of annealing.
Relative to the pre-irradiation level of less than 10 nA, the current 
at a bias voltage of 600 V is seen to rise
to a fraction of a microamp after irradiation, leading to leakage 
currents of less than 1 $\mu$A/cm$^2$ over the 0.21 cm$^2$ area of the
GaAs pad sensors. Annealing is seen
to markedly reduce the post-irradiation leakage current, although as discussed
above, the same annealing steps dramatically reduced the CCE.

Table~\ref{tab:cc_results} provides a table of maximum median collected
charge, both before and after irradiation, and median
charge loss due to irradiation, for all sensors studied.

\begin{table*}[h]
\begin{centering}
\caption{Maximum median CCE before and
after irradiation. While the NC02
sensor was irradiated at a temperature of 5$^o$ C,
it was accidentally annealed for approximately 5 hours
at temperatures as high as 130 C. Room-temperature 
annealing was required to recover the full CC for the 
NF07 sensor. For the WSI-P4 sensor (PF pad sensor),
the quoted charge CCE is that after
annealing at 60$^o$ C For the GaAs sensors, the quoted
CCE is that before the
sensor spent a significant amount of time above 
-10$^o$ C; annealing was found to significantly degrade the CCE
for GaAs sensors.
}
\label{tab:cc_results}
\vspace {5mm}
\begin{tabular}{lcccc}
Sensor    & Dose    & Median CC Before  & Median CC After      &  Fractional      \\
         &   (Mrad) &     Irradiation (fC) & Irradiation (fC)         & Loss (\%)    \\ \hline \hline
PF05   & 5.1  &  3.70  &  3.43  &   7     \\
PF14   & 20   &  3.68  &  3.01  &  18     \\
PC08   & 20   &  3.51  &  3.09  &  12     \\
NF01   & 3.7  &  3.76  &  3.81  &   0     \\
NF02   & 19   &  3.75  &  3.60  &   4     \\
NF07   & 91   &  3.75  &  4.00  & 0       \\
NC01   & 5.1  &  3.71  &  3.80  &   0     \\
NC10   & 18   &  3.76  &  3.74  &   1     \\
NC03   & 90   &  3.68  &  3.55  &   4     \\
NC02   & 220  &  3.69  &  3.06  &  17     \\  \hline
WSI-P4 (PF) ($@600$ V) & 269 & 3.77 & 3.17 &  16    \\ \hline
GaAs18 ($@600$ V)      & 5.7  &  6.41 & 5.22 & 19 \\ 
GaAs09 ($@600$ V)      & 20.8 & 4.74  & 2.02 & 57 \\ \hline

\end{tabular}
\par\end{centering}
\end{table*}

\section{Summary and Conclusions}

We have explored the radiation tolerance of four different types of
silicon diode sensors (n-type and p-type float zone and magnetic Czochralski bulk) sensors, 
as well as a GaAs sensor,
exposing them to doses as high as 270 Mrad
at the approximate maxima of tungsten-induced electromagnetic showers.
For the two types of silicon diode sensors (n-type Czochralski and
p-type float zone) explored with several-hundred Mrad doses,
we have found charge-collection efficiency in excess of 80\% for a bias voltage of 600 V.
For the p-type float zone sensor after irradiation with 270 Mrad, the leakage current
at a temperature of -10$^o$ C was measured to be approximately 80 $\mu$A/cm$^2$. 
This suggests that, with the application of active cooling, silicon diode sensors may be
sufficient for the operation of a calorimeter exposed to
hundreds of Mrad, approaching the specification required for
the most heavily irradiated sensors in the ILC BeamCal instrument. 
While the depletion voltage increases significantly with
radiation, studies of the charge-collection efficiency as a function of annealing temperature
suggests that 
the depletion voltage can be significantly decreased after one-hour annealing runs
at temperatures up to 80$^o$ C. 
On the other hand, the GaAs sensors showed significant loss of charge-collection efficiency
even for modest radiation doses of 6 Mrad; for a dose of 21 Mrad the charge-collection efficiency
was reduced by nearly 60\%. Room-temperature annealing further reduced the charge-collection
efficiency to 17\% of the value measured before irradiation. In all cases, however, the leakage current
remained below 1 $\mu$A/cm$^2$. 

In the most recent run of experiment T-506, a 
full complement of n- and p-type float zone and magnetic Czochralski pad sensors 
were exposed to doses of approximately 300 Mrad each. For the case of the
p-type float zone sensor, which had already been irradiated and studied as discussed
above, the total accumulated dose is now approaching 600 Mrad. In addition, a silicon carbide
sensor was exposed to a dose of 100 Mrad. The CCE and leakage current performance of
these sensors is currently under evaluation.

\section{Acknowledgments}

We are grateful to Leszek Zawiejski, INP, Krakow for supplying us with the tungsten plates
needed to form our radiator, and Georgy Shelkov, JINR, Dubna for supplying us with GaAs
sensors for irradiation studies.  We would also like to express our gratitude
to the SLAC Laboratory, and particularly the End Station Test Beam delivery
and support personnel, who made the run possible and successful.
Finally, we would like to thank our SCIPP colleague Hartmut Sadrozinski for
the numerous helpful discussions and guidance he provided us.

\section{Role of the Funding Source}

The work described in this article was supported by the United States Department of Energy,
DOE contract DE-AC02-7600515 (SLAC) and grant DE-FG02-04ER41286 (UCSC/SCIPP). The funding agency
played no role in the design, execution, interpretation, or
documentation of the work described herein.




\nocite{*}
\bibliographystyle{elsarticle-num}
\bibliography{martin}



\end{document}